\documentclass{pasj00}
\color{red}
\color{black}
\SetRunningHead{S. Konami et al.}{
Abundance Patterns in the Outflow Region of M~82 and the Importance of CX}

\title{Suzaku Metal Abundance Patterns in the Outflow Region of
M82 and the Importance of Charge Exchange}
\author{
 Saori \textsc{Konami},\altaffilmark{1,2}
 Kyoko \textsc{Matsushita},\altaffilmark{1} 
 Takeshi Go \textsc{Tsuru},\altaffilmark{3}
 Poshak \textsc{Gandhi},\altaffilmark{4} \\
and Toru \textsc{Tamagawa}\altaffilmark{2,1}
}
\altaffiltext{1}{
Department of Physics, Tokyo University of Science,
1-3 Kagurazaka, Shinjuku-ku, Tokyo 162-8601}
\altaffiltext{2}{
High Energy Astrophysics Laboratory, the Institute of Physical and Chemical Research,
2-1 Hirosawa, Wako, Saitama 351-0198}
\email{konami@crab.riken.jp}
\altaffiltext{3}{
Department of Physics, Kyoto University, Kitashirakawa-oiwake-cho, 
Sakyo-ku, Kyoto 606-8502}
\altaffiltext{4}{
Institute of Space and Astronautical Science (ISAS), Japan Aerospace Exploration Agency, 
3-1-1 Yoshinodai, Chuo-ku, Sagamihara, Kanagawa 252-5210}

\Received{}
\Accepted{}
\Published{---}

\begin{document}
\maketitle
\begin{abstract}

We performed spectral analysis of Suzaku data of the galactic disk and outflow regions of
the starburst galaxy M82. Thermal modeling of the central disk regions requires 
at least three temperature components. The Ly$\beta$ line fluxes of
O \emissiontype{VIII} and Ne \emissiontype{X} exceed those expected from a plasma in collisional ionization equilibrium. 
The ratios of Ly$\beta$/Ly$\alpha$ lines for O \emissiontype{VIII} and Ne \emissiontype{X} 
are higher than those of collisional ionization equilibrium, which may be caused by the process of charge exchange. 
In the outflow wind region, the spectra are
well reproduced with two-temperature thermal models, 
and we have derived the metal abundances of O, Ne, Mg, and Fe in the outflow.
The ratios of O/Fe, Ne/Fe, and Mg/Fe are about 2, 3, and 2, respectively,
relative to the solar value determined by Lodders (2003). 
Since there is no evidence of charge exchange in outflow region, 
the metal abundances should be more reliable than those in the central region.
This abundance pattern indicates that starburst activity enriches the
outflow through SN II metal ejection into intergalactic space.

\end{abstract}

\section{Introduction}

Metals in hot gas in the interstellar medium (ISM) and intergalactic medium
can provide a cumulative fossil record on the history of star formation and
evolution of galaxies.
Starburst galaxies play an important role in supplying intergalactic metals
via outflows heated by type II supernovae (SN II).
These outflows are thought to contain metals synthesized
by SN II and may have enriched the intra-cluster medium (ICM)
and warm-hot intergalactic medium (WHIM).

Recent X-ray observations of starburst galaxies have revealed extended halos of hot plasma around several galaxies 
(e.g. \cite{wang_05}; \cite{strick_07b}; \cite{wang_09}).
However, measurements of metal abundances within these outflows are limited. 
For NGC~253, XMM-Newton detected the metals O, Ne, Mg, Si, and Fe 
in the central regions $\lesssim$3~kpc along the outflow axis (e.g. \cite{bauer_07}). 
For NGC~4631 the abundance patterns for O, Ne, Mg, Si and Fe were observed in the extended 
halo region, $\sim$13~kpc with Suzaku \citep{yamasaki_09}. 
The outflow wind and disk regions of NGC~4631 show patterns close to those expected from the SN II yields 
and solar abundance (\cite{lodders_03}), respectively.

M~82 is the most famous, prototypical starburst galaxy with
an X-ray emitting outflow  (e.g. \cite{tsuru_97}; \cite{stevens_03}; \cite{origlia_04}). 
\citet{tsuru_97} observed M 82 with ASCA and derived a peculiar abundance pattern 
of O/Ne/Mg/Si/S and Fe in the hot X-ray emitting plasma around the starburst region. 
No combination of SN Ia and SN II could reproduce the observed abundance pattern. 
Furthermore, \citet{ranalli_08} reported spatial variations of
the abundances with XMM-Newton. 
They found the abundances of O, Ne, Mg, Si, and Fe 
to increase from the center going out to the external outflow region, 
the absolute value of the O abundance they found was extremely low. 
In contrast, 
\citet{tsuru_07} derived SN II like abundance pattern in the `cap' region,
which may be a large photoionized cloud in the halo of M82,
at a distance of 11.6~kpc north from the center.

\citet{lallement_04} has pointed out that line emission in the `cap' region 
could arise from the process of charge-exchange (CX), where highly ionized gas may be colliding with the cold cloud. 
\citet{tsuru_07} reported that an emission line consistent with the
C \emissiontype{VI} transition of n = 4 to 1 at 0.459 keV has been marginally
detected, although it is not statistically significant at
the 99\% confidence level; the presence of this line would suggest
charge-exchange processes in the `cap' region with Suzaku.
In the XMM spectra in the central region, \citet{ranalli_08} 
reported line-like features around $\sim$ 0.7~keV
and 1.2~keV, which were interpreted as CX reactions of neutral Mg and Si, respectively. 
Most recently, in the XMM reflection grating spectra (RGS),
 \citet{liu_11} reported detection of the CX emission of O \emissiontype{VII},  Ne \emissiontype{IX}, and Mg \emissiontype{XI} triplets at the center region.
An important piece of evidence of CX is that 
the forbidden lines are stronger than the resonance lines, 
especially for O \emissiontype{VII} ions.
They concluded that a significant fraction of these lines 
originate as a result of CX emission rather than thermal radiation.

CX occurs between highly ionized atoms and neutrals, and emits X-ray emission lines. 
When a highly ionized ion hits a neutral atom, 
an electron is captured in an excited state in the ion.
It then decays to the ground state, emitting X-ray photons. 
CX emission has been detected from regions where hot and cool gas
co-exist. 
The first discovery of CX was from comet Hyakutake \citep{lisse_96}, following 
which X-ray emission was detected from many comets. More recently, the diffuse 
soft X-ray background was found to contain CX emission from neutrals in the geocorana/heliosphere 
and solar wind ions (e.g. \cite{cox_98}; \cite{cravens_00}).
In studies of supernova remnants, CX emission can explain 
anomalously high intensity ratios (i.e. $\beta$/$\alpha$ and $\gamma$/$\alpha$ lines) 
compared with those of collisional ionization equilibrium (CIE) plasma (\cite{rasmussen_01}, \cite{katsuda_11}). 
In the M~31 bulge, an intensity excess at the O \emissiontype{VII} triplet was
observed owing to CX emission \citep{liu_10}.

Over the 3~keV energy range in the spectra of M~82, 
there are hard components whose nature is not yet fully understood. 
Some part of this emission comes from point sources (X-1, for example;
\cite{miyawaki_09}). 
With Chandra and XMM-Newton,
\citet{strick_07} detected diffuse hard X-ray emission and Fe-K line emission
which cannot be attributed to CIE plasma.

In this paper, we investigated the spatial distribution of abundance patterns in the 
outflow wind of M~82, between the center and the `cap' regions.
We also derived ratios of K$\beta$ to K$\alpha$ lines of O and Ne
and studied the spatial variation of the effect of CX emission.
We expect both hot ions and cold atoms exist in the center and the `cap' regions. 
This paper is structured as follows. 
In section 2, we summarize observations of Suzaku and 
XMM-Newton. 
Sections 3 and 4 detail the data analysis and results. 
Section 5 gives a discussion of these results. 
Finally, in section 6 we present our conclusions. 

We adopted 3.53~Mpc for 
the distance to M~82, which was determined by the luminosity 
of the tip of red giant branch \citep{karachentsev_04}. 
Unless noted otherwise, we use the new solar abundances in \citet{lodders_03}, 
and the quoted errors are for a 90\% confidence interval 
for a single interesting parameter.

\section{Observation and Data Reduction}\label{sec:obs}
\subsection{Suzaku}

Suzaku observed M~82 on three occasions in 2005 October during 
the Science Working Group phase to investigate the outflowing 
hot gas from the galactic plane.
The logs of three observations are listed in table\ref{table1}. 
The XIS consists of three front-illuminated (FI: XIS0, XIS2 and XIS3) CCD cameras 
and one back-illuminated (BI: XIS1) CCD camera \citep{koyama_07}. 
Combined with XRT \citep{serle_07}, the field of view (FOV) of the 
XIS covers a $\sim$18$'$$\times$18$'$ region with a half-power diameter (HPD) 
of $\sim$2$'$.
The XIS was operated in normal 
clocking mode (8~s exposure per frame), with the standard 
5 $\times$ 5 and 3 $\times$ 3 editing mode. 
We processed the XIS data using the ``xispi'' and ``makepi'' ftool tasks 
and CALDB files of version 2009-08-13. 
The XIS data were then cleaned by assuming thresholds on the Earth 
elevation angle of $>~5^{\circ}$ and the Day-Earth elevation angle of 
$>~20^{\circ}$. 
We also discarded data with time since south Atlantic anomaly passage of less than 436 sec. 
After this screening, the remaining good exposure was 101.1 ksec 
for both FIs and BI summed over the three observations.
Event screening with cut-off rigidity was not performed.

The spectral analysis was performed with HEAsoft version 6.10 and XSPEC 12.6.
To subtract the non-X-ray background (NXB), we employed 
the Dark-Earth database using the ``xisnxbgen'' ftool task \citep{tawa_08}.
We generated two different ancillary response files (ARFs) 
for the spectrum of each region; one assumed uniform 
sky emission, while the other utilized the observed XIS1 image by the 
``xissimarfgen'' ftool task \citep{ishisaki_07}. In the ARFs, we also 
included the effect of contamination on the optical blocking filters (OBF) of the XIS.

\subsection{XMM-Newton}
As mentioned in \citet{tsuru_07}, since there is a luminous point source within the outflow of M~82, 
we analyzed the XMM-Newton data to estimate its contamination.
Although XMM-Newton observed M~82 four times, we 
analyzed the data from 2004 April. This observation has the single largest 
exposure time, which is sufficient for this analysis.
The log of this observation is given in table \ref{table1}.
The European Photon Imaging Camera (EPIC), aboard XMM-Newton, has two metal 
oxide semiconductor (MOS) CCD arrays and one PN CCD array. All three cameras 
have a $\sim$30$'$ diameter circle FOV with a HPD of $\sim$6$''$, which is much smaller 
than the XRT and hence advantageous for observations of point sources. 

All data were processed using version 10.0.0 of the XMM-Newton Science 
Analysis Software (SAS). Periods with a 10--15 keV background count rate deviating by more than 3$\sigma$ from the mean rate were filtered out. 
The resultant exposure times are also listed in table \ref{table1}.

\begin{table*}
\caption{
Logs of Suzaku and XMM-Newton observations
}
\label{table1}
\begin{center}
\begin{tabular}{lccc} \hline\hline
Instrument   &  Observation   &  Observation   & Effective   \\ 
             &      ID         &  Start Date    & Exposure (ks) \\ \hline
XIS          & 100033010          & 2005/10/04               & 32.3   \\
             & 100033020          & 2005/10/19               & 40.4   \\
             & 100033030          & 2005/10/27               & 28.4   \\
EPIC         & 0206080101         & 2004/04/21               & 66.8/68.2/52.9$^{\ast}$  \\ \hline \hline
\end{tabular}
\end{center}
\parbox{\textwidth}{\footnotesize
\footnotemark[$\ast$] 
The exposure times of the MOS1/MOS2/PN.
}
\end{table*}

\section{Analysis and Results for the Disk region}
\subsection{Extraction Regions of Spectra}
\label{region}
\begin{figure*}
\begin{center}
\centerline{
\FigureFile(\textwidth,\textwidth){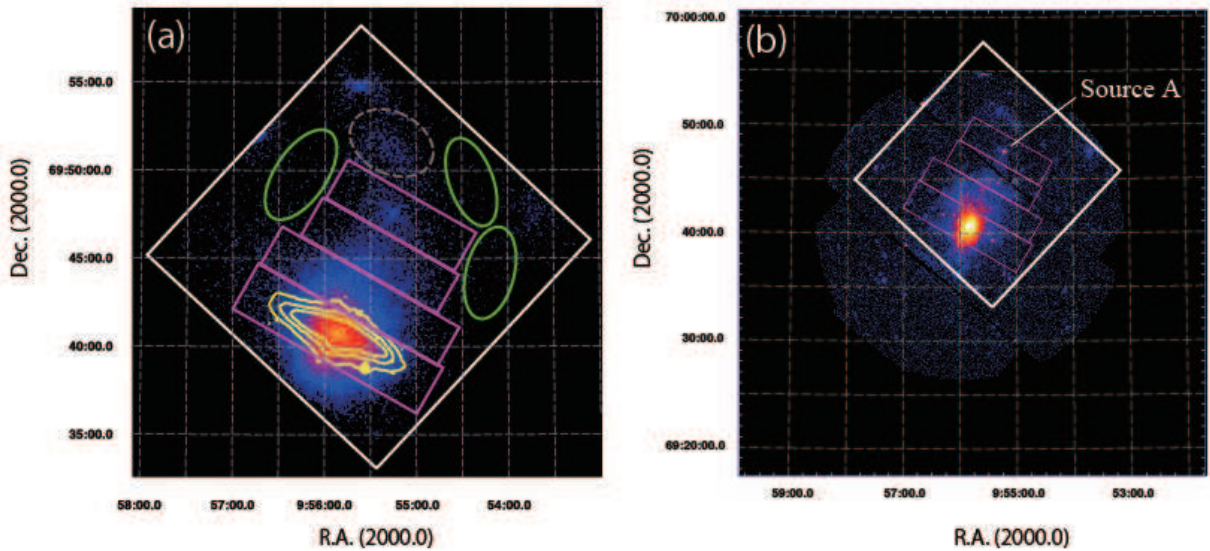}}
\caption{
Panel (a) and (b) are the X-ray images of M~82 with the Suzaku XIS1 
and XMM-Newton EPIC-MOS1, respectively. 
The background components were not subtracted and vignetting 
was not corrected for in these images.
Panel (a); the XIS~1 image in 0.3--2 keV is smoothed with a $\sigma$=5.2$''$ Gaussian profile and 
overlaid with optical galactic disk contours from the Digitized Sky Survey (yellow). 
The magenta boxes show the regions for spectra extraction labeled Disk, Wind-1, 
Wind-2, and Wind-3 going out from the center. 
The background spectra were extracted from the three green ellipses.
The gray ellipse shows the `cap' region \citep{tsuru_07}.  
Panel (b); the selected Wind regions are overlaid. The position of Source~A 
is also indicated. The white box is the Suzaku XIS field of view in both panels. 
}\label{figure1}
\end{center}
\end{figure*}

Figure \ref{figure1} (a) shows the Suzaku X-ray image in energy range of 0.3--2~keV range, along with the 
spatial regions used for spectral analysis.
We defined the square region within 3$'$$\times$12$'$ of the center of M~82 as 
Disk region.
Moreover, because there may be spatial variations in abundance patterns \citep{ranalli_08}, the region between Disk and `cap' were divided 
into three regions as shown in figure \ref{figure1} (a). 
These three regions are labeled Wind-1, Wind-2, and Wind-3 in order, 
from the center.
In this paper, we derived abundance patterns of these regions. 
The three regions outside the M~82 wind are selected to extract background emission, and are shown 
as green ellipses (figure {\ref{figure1}} (a)). 
We added the spectra of these three regions in order to derive the background spectra.

In the figure \ref{figure1} (b), we show the XMM-Newton EPIC image in 
0.3--5~keV range, 
overlaid with the extraction regions for Suzaku.
As mentioned in \citet{tsuru_07}, the luminous X-ray source in the outflow 
lies in the region Wind-3, and we refer to it as Source~A, hereafter.

\subsection{Fitting the Spectra}
\label{disk spectra}

We fitted the Disk region spectra with following model: 
phabs$_{\rm M82}$ $\times$ vapec$_{\rm 1T,~2T,~or~3T}$ + phabs$_{\rm X-1}$ $\times$ cutoff PL$_{\rm X-1}$. 
The phabs$_{\rm M82}$ and phabs$_{\rm X-1}$ factors represent the column density. The first of these is left free 
and the second one is fixed to 1.1 $\times$ 10$^{22}$ cm$^{-2}$ \citep{miyawaki_09}. 
The ISM emission of M~82, vapec$_{\rm 1T,~2T,~or~3T}$, was represented by 
one-temperature (1T), two-temperature (2T), and three-temperature (3T) models,
employing the vapec plasma code \citep{smith_01}. 
We divided the metals into the groups of O, Ne, (Mg \& Al), Si, S, (Ar \& Ca) and (Fe \& Ni) 
based on the metal synthesis mechanism of SN, and allowed each group to vary.
In the 2T or 3T modeling, the vapec components were assumed to share the 
same elemental abundances. 
Because our purpose is to derive metal abundance of CIE hot plasma, 
we accepted the best fit absorbed cutoff power-law (PL) model from \citet{miyawaki_09} 
for the hard component, which is dominated by emission from X-1.
The cutoff energy is fixed to 5.8~keV, which is 
the best-fit value in \citet{miyawaki_09}. 
Meanwhile, the photon index and normalization are set free.  
We simultaneously fitted the spectra of all CCDs, 
using the energy ranges of 0.4--6 keV for the BI CCD and 
0.5--6 keV for the FI CCDs. 
Because there are known calibration issues of the Si edges in all CCDs, 
the energy range of 1.84--1.86~keV is ignored, hereafter \citep{koyama_07}.

\begin{figure*}
\begin{center}
\centerline{
\FigureFile(0.6\textwidth,0.6\textwidth){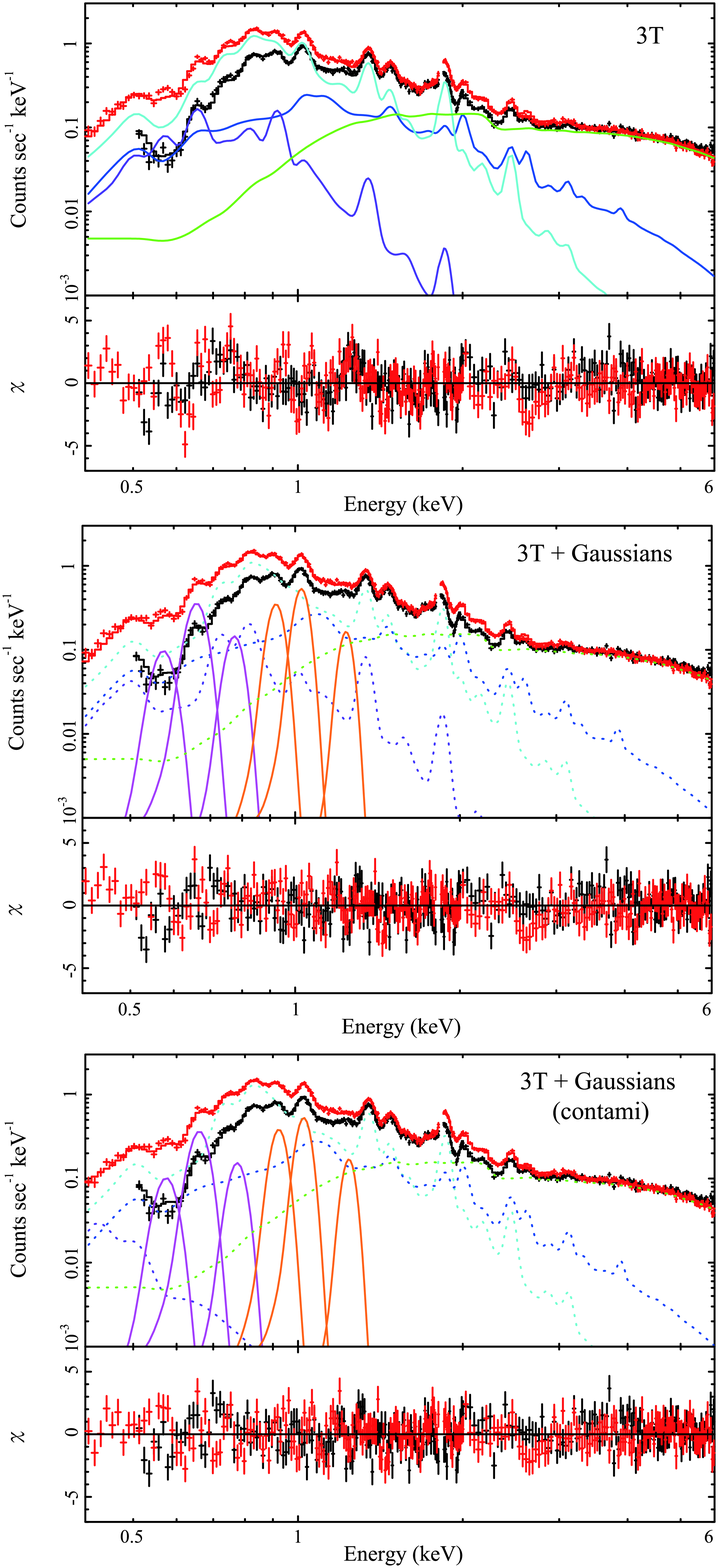}}
\caption{
Background-subtracted XIS0 (black) and XIS1 (red) spectra of Disk, shown 
without removal of the instrumental response. 
In order from top to bottom, the panels are the spectra fitted with model 3T for ISM, 3T + Gaussian lines, 
and 3T + Gaussian lines with absorption of OBF contaminant. 
Black and red lines show the best-fit model in each case for FIs and XIS1, respectively. 
For simplicity, only the model components plot of the XIS1 spectra are shown. 
Light blue, blue and purple lines are the three ISM temperature components. 
Green curves represent the accumulated emission from point sources. 
Magenta and orange lines are the Gaussians for O \emissiontype{\rm VII/VIII} and 
Ne \emissiontype{\rm IX/X}, respectively. 
}\label{figure2}
\end{center}
\end{figure*}

The 1T, 2T, and 3T models failed to reproduce the spectra, 
${\chi^2}$/d.o.f.=2376/942, 1784/940, and 1673/938 respectively, despite improving fit statistic values. 
There are large residuals around 0.5--0.7 and $\sim$1.2~keV as shown in top of figure \ref{figure2}.
Although we tested fits with 4T and 5T models, the residuals features and fit statistics do not change significantly, 
${\chi^2}$/d.o.f.=1647/936 and 1653/934, respectively. 
Therefore, we treated thermal components represented by 3T models. 
The large residuals existing around 0.5--0.7 and $\sim$1.2~keV correspond to 
O  \emissiontype{VII/VIII} and Ne \emissiontype{X} Ly$\beta$ emission lines, respectively. 
We considered that these highly ionized O and Ne may occur CX interacting with neutrals.
Adding six gaussians to represent O and Ne emission lines at 0.574~keV (O \emissiontype{VII} K$\alpha$), 
0.66~keV (O \emissiontype{VII} K$\beta$ + O \emissiontype{VIII} Ly$\alpha$), 
0.774~keV (O \emissiontype{VIII} Ly$\beta$), 0.921~keV (Ne \emissiontype{IX} K$\alpha$), 
$\sim$1.02~keV (Ne \emissiontype{IX} K$\beta$ + Ne \emissiontype{X} Ly$\alpha$), 
and $\sim$1.2~keV (Ne \emissiontype{X} Ly$\beta$), 
we fitted the spectra with the above model while setting O and Ne abundance of vapec models to zero.
The Gaussian line centers for only 1.02~keV and 1.2~keV are set free, 
because relatively strong Fe-L line emission exists in this energy ranges. 
The phabs component is fixed at 29.9 $\times$ 10$^{20}$ cm$^{-2}$, which is value of fitting with 3T models.  
This model produced a best fit with ${\chi^2}$/d.o.f.=1364/933 and the derived parameters 
are listed in table \ref{table2}. 

Although the large residuals disappear with the above fit, 
the discrepancies of residuals between XIS0 and XIS1 still remain, over the energy range of 0.5--0.7~keV. 
These residuals were attributed to contamination on the optical blocking filters (OBF) on the XIS.
It is known that contaminant has been accumulating on the OBF since the detector doors 
were opened following launch \citep{koyama_07}.  
To estimate the effect of absorption, we added the model `varabs', which treats 
photoelectric absorption with variable abundances. 
The parameters of C is set free. 
Meanwhile, we set O to 1/12 of C in units of the solar ratio, 
assuming the contaminant material of the OBF to be C$_{24}$H$_{38}$O$_{4}$ 
and owing to the solar units of the abundances in the model. 
We fitted the spectra with a model of 3T plus gaussians multiplied varabs, which yielded 
${\chi^2}$/d.o.f. of 1293/929. The fitted spectra are shown in the bottom panel of figure \ref{figure2}.
The column density of C and O are 2.2--7.8 and 0.37--1.33 $\times$ 10$^{17}$ cm$^{-2}$, respectively. 
These values are consistent with those during other early observations \citep{fujimoto_07}. 
The derived parameters are listed in table \ref{table2}. 
Although the fits are not formally acceptable, these results are useful for assessing whether CX is important or not.

\begin{table*}
\caption{
Summary of the best-fit parameters for the Disk region with 3T + Gaussian models.}
\label{table2}
\begin{center}
\begin{tabular}{lccccccc} \hline \hline
Parameters & & Disk & & &      \\ 
           &  &  3T for ISM & &         \\
           &  &  &Gaussians  & Gaussians            \\
           &   &  &     &  (contami)   \\\hline
$N_{\rm H_{M82}}$ & (10$^{20}$ cm$^{-2}$) & 29.9$^{+1.22}_{-1.20}$ & 29.9 (fix)  & 29.9 (fix) \\
$kT_{\rm 1T}$ & (keV) & 1.93$^{+0.11}_{-0.12}$ & 1.70$^{+0.03}_{-0.08}$  & 1.62$^{+0.04}_{-0.08}$  \\ 
$kT_{\rm 2T}$ & (keV) & 0.59$\pm0.01$ & 0.61$\pm0.01$  & 0.58$\pm0.01$    \\ 
$kT_{\rm 3T}$ & (keV) & 0.24$\pm0.02$ & 0.34$^{+0.04}_{-0.02}$  & 0.08$\pm0.01$    \\ 
O & (solar) & 0.45$^{+0.05}_{-0.04}$ & 0 (fix)  & 0 (fix)   \\
Ne & (solar) & 1.13$^{+0.09}_{-0.08}$ & 0 (fix) & 0 (fix)   \\
Mg, Al & (solar) & 1.04$\pm0.08$ & 1.13$^{+0.09}_{-0.05}$ & 1.09$^{+0.06}_{-0.11}$   \\
Si & (solar) & 1.23$^{+0.09}_{-0.08}$ & 1.23$^{+0.09}_{-0.04}$ & 1.18$^{+0.03}_{-0.05}$   \\
S & (solar) & 1.52$\pm0.11$ & 1.40$^{+0.08}_{-0.08}$ & 1.38$^{+0.07}_{-0.08}$   \\
Ar, Ca & (solar) & 0.65$\pm0.31$ & 1.36$^{+0.27}_{-0.26}$ & 1.38$^{+0.32}_{-0.26}$   \\
Fe, Ni & (solar) & 0.38$\pm0.03$ & 0.42$^{+0.05}_{-0.01}$  & 0.42$^{+0.03}_{-0.04}$     \\
$\Gamma_{\rm cutoff~PL}$  &  & 0.47$^{+0.01}_{-0.03}$ & 0.52$\pm0.03$ & 0.55$^{+0.02}_{-0.04}$   \\ [1.0ex]
line energy & (keV) &-& 0.574 (fix) & 0.574 (fix)   \\
Norm & (10$^{-4}$) &-& 1.28$\pm0.16$ & 1.35$^{+0.16}_{-0.15}$ \\
line energy & (keV) & -& 0.66 (fix) & 0.66 (fix)  \\
Norm & (10$^{-4}$) & -&3.34$\pm0.13$ & 3.45$^{+0.14}_{-0.13}$ \\
line energy & (keV) & -& 0.774 (fix) & 0.774 (fix)  \\
Norm & (10$^{-4}$) & -& 1.08$\pm0.12$  & 1.16$\pm0.12$ \\
line energy & (keV) &- & 0.921 (fix) & 0.921 (fix) \\
Norm & (10$^{-4}$) &- & 2.24$\pm0.10$ & 6.52$^{+0.20}_{-0.10}$ \\
line energy & (keV) &- & 1.03$\pm0.01$ & 1.02$\pm0.01$ \\
Norm & (10$^{-4}$) &- & 3.20$\pm0.09$ & 3.22$\pm0.09$ \\
line energy & (keV) &- & 1.24$\pm0.01$ & 1.24$\pm0.01$ \\
Norm & (10$^{-4}$) &- & 0.95$\pm0.05$ & 0.99$\pm0.05$ \\[2.0ex]
${\chi^2}$/d.o.f. & & 1673/938 & 1364/933 & 1293/929  \\ \hline \hline
\end{tabular}
\end{center} 
\end{table*}

\section{Analysis and Results for the Wind regions}

\subsection{Contamination of Disk Emission}
\label{contami}
\begin{figure*}
\begin{center}
\centerline{
\FigureFile(0.6\textwidth,0.6\textwidth){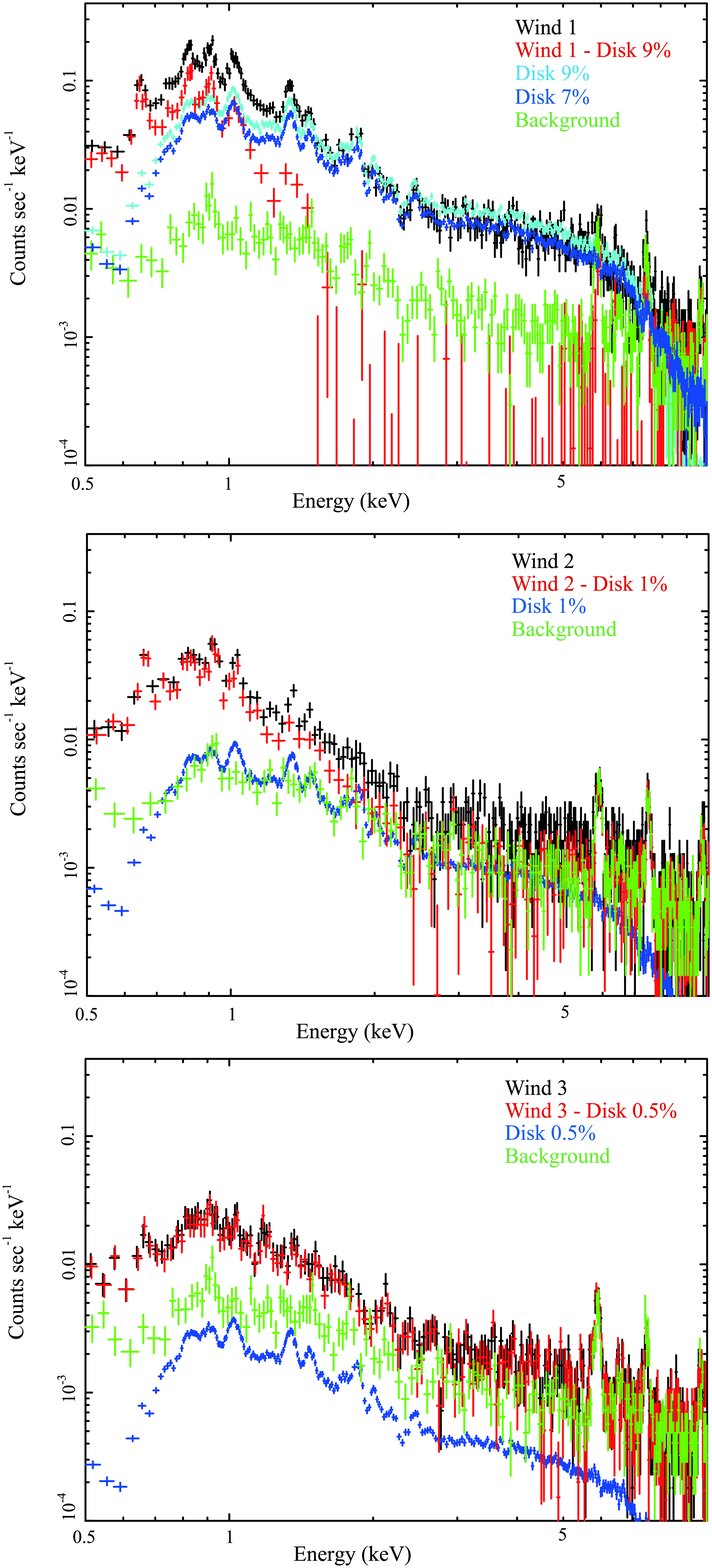}}
\caption{
From top to bottom, panels are the comparison spectra between scaled contamination Disk and 
Wind-1, Wind-2, and Wind-3, respectively.
In the top panel, the spectra of Wind-1 and Disk scaled by 9\% and 7\% are shown in black, light blue, and 
blue, respectively.
The contamination-subtracted Wind-1 and background region spectra are shown as red and green, respectively. 
The other panels follow the same layout. 
}\label{figure3}
\end{center}
\end{figure*}

The Disk emission contaminates the Wind regions owing to the low Suzaku 
angular resolution (HPD $\sim$ 2$'$).
We simulated the emission from the Disk region in each Wind region, setting the point source to 
the coordinates of X-1 by xissimarfgen 
in the energy range of 0.2--16~keV, 
because the Suzaku PSF does not allow us to resolve the X-ray peak of hot ISM and 
X-1 \citep{ishisaki_07}. 
We found a level of 9\%, 1\%, and 0.5\% Disk intensity contamination to  
the Wind-1, Wind-2, and Wind-3 regions, respectively.
The level of contamination is almost constant, to within 5\% below 6 keV,
and within 20\% at 6--10 keV.
As shown in figure \ref{figure3}, most of the Wind-1 emission beyond 1.5~keV can be attributed to the  
Disk region. Meanwhile, in the Wind-2 and Wind-3 regions, the contamination from 
Disk emission is smaller compared with Wind-1. 
The intensity of contamination from Disk region is one-half and one-fifth 
of that in Wind-2 and Wind-3 above $\gtrsim$1.5 keV, respectively. 
Hereafter, we analyze the spectra of Wind-1 and Wind-2 regions
after subtraction of the directly-scaled emission of the Disk region. 
Full statistical error propagation was employed in this analysis.

\subsection{Wind-1 Region}
\label{halo1}

We first fitted the Wind-1 region spectra after subtracting 9\% of the Disk emission 
and the background emission accumulated over three background regions
with the following model;
phabs$_{\rm G}$ $\times$ vapec$_{\rm 2T}$. 
The Wind-1 region spectra will be reproduced as the 
sum of Disk emission, power-law component of X-1, and Wind-1 emission. 
The phabs$_{\rm G}$ factor is fixed to 
4.0 $\times$ 10$^{20}$ cm$^{-2}$, which is the Galactic value of $N_{\rm H}$ in the direction of M~82. 
The abundances of O, Ne, Mg, and Fe are allowed to vary, and those of other metals are fixed to solar value.  
We simultaneously fitted the spectra of all CCDs over the 
energy ranges of 0.4--5.0 keV for the BI CCD and 
0.5--5.0 keV for the FI CCDs.
The 1T and 2T models failed to reproduce the spectra, with ${\chi^2}$/d.o.f.=845/305 and 807/303, 
respectively. 
Above 2~keV, there are large negative residuals which we attribute 
to excess subtraction of the Disk emission contamination. 

\begin{figure}
\begin{center}
\centerline{
\FigureFile(0.5\textwidth,0.5\textwidth){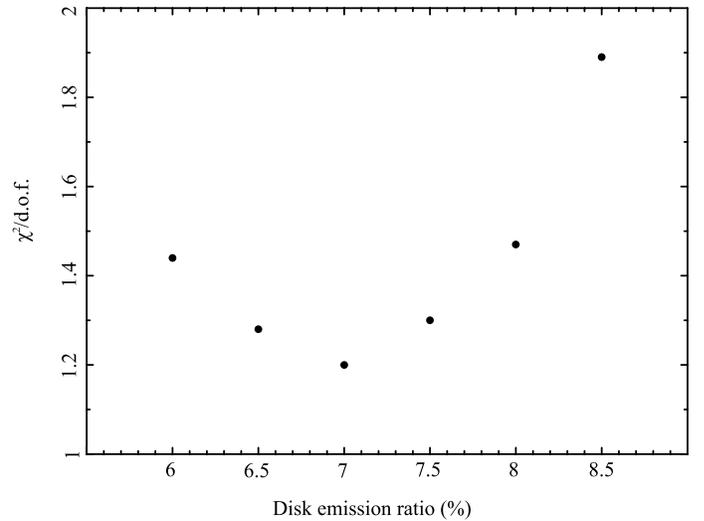}}
\caption{
The ${\chi^2}$/d.o.f. distribution obtained when fitting the Wind-1 spectra, after subtraction of  various Disk contamination levels.
}\label{figure4}
\end{center}
\end{figure}

\begin{figure*}
\begin{center}
\centerline{
\FigureFile(\textwidth,\textwidth){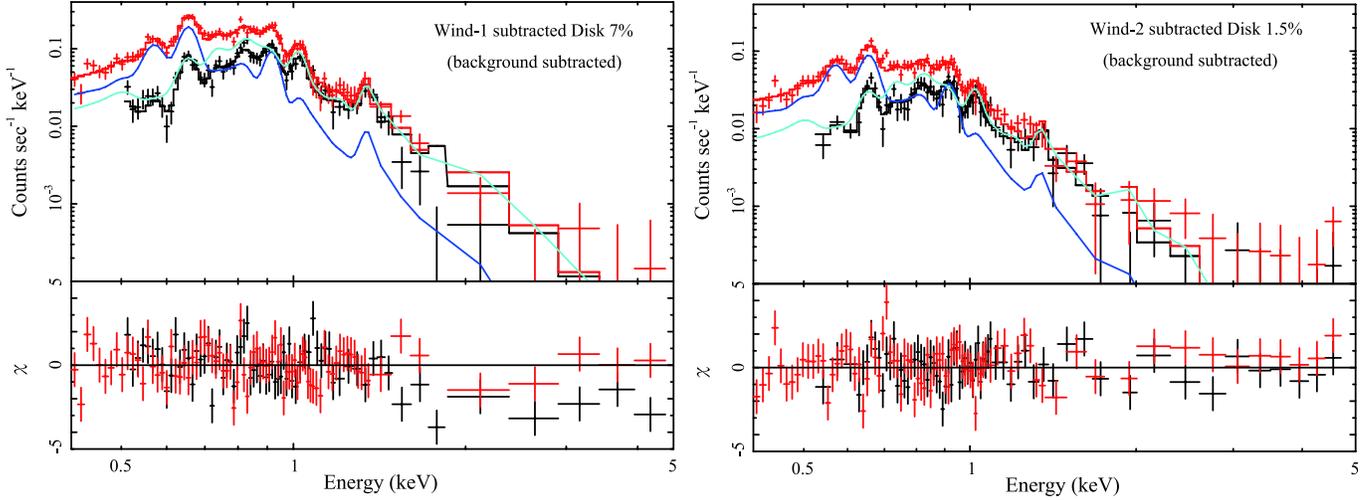}}
\caption{
Background-subtracted XIS0 (black) and XIS1 (red) spectra 
of Wind-1 (left) and Wind-2 (right), shown without removal of instrumental response. 
Black and red lines show the best-fit model for the XIS0 and XIS1, 
respectively. For simplicity, only the model components for XIS1 spectra 
are shown. Blue and light blue lines are the ISM components. 
}\label{figure5}
\end{center}
\end{figure*}

The simulated intensity of Disk contamination may have some
uncertainties due to spatial spread of the X-ray emission
and uncertainties in the point spread function.
Therefore, we tested the effect of varying the Disk contamination level over 6--8.5\%, subtracting off this component, 
and then carrying out the fit.   
The ${\chi^2}$/d.o.f. values are plotted in figure \ref{figure4}, and are found to show a minimum at 354/294 with 
the contaminant level of 7\% of the Disk emission.
Above 2 keV, the spectra of the Wind-1 region are well reproduced by
the contamination of 7\% emission from the Disk region (figure \ref{figure3}), and these parameters were adopted hereafter. 
The derived parameters and fitted spectra are shown in the first row of 
table \ref{table3} and in figure \ref{figure5}, respectively.
As shown in figure \ref{figure5}, in the Wind-1 spectra, several emission lines are 
detected. The ones around 0.5--0.6 keV, 0.6--0.7 keV and $\sim$1.3 keV are identified with 
K$\alpha$ lines of O \emissiontype{VII}, O \emissiontype{VIII}, and 
Mg \emissiontype{XI}, respectively.
The emission bump around 0.7--1 keV corresponds to 
Fe-L complex, as well as to K-lines from Ne \emissiontype{IX} and Ne \emissiontype{X}. 

\begin{figure*}
\begin{center}
\centerline{
\FigureFile(\textwidth,\textwidth){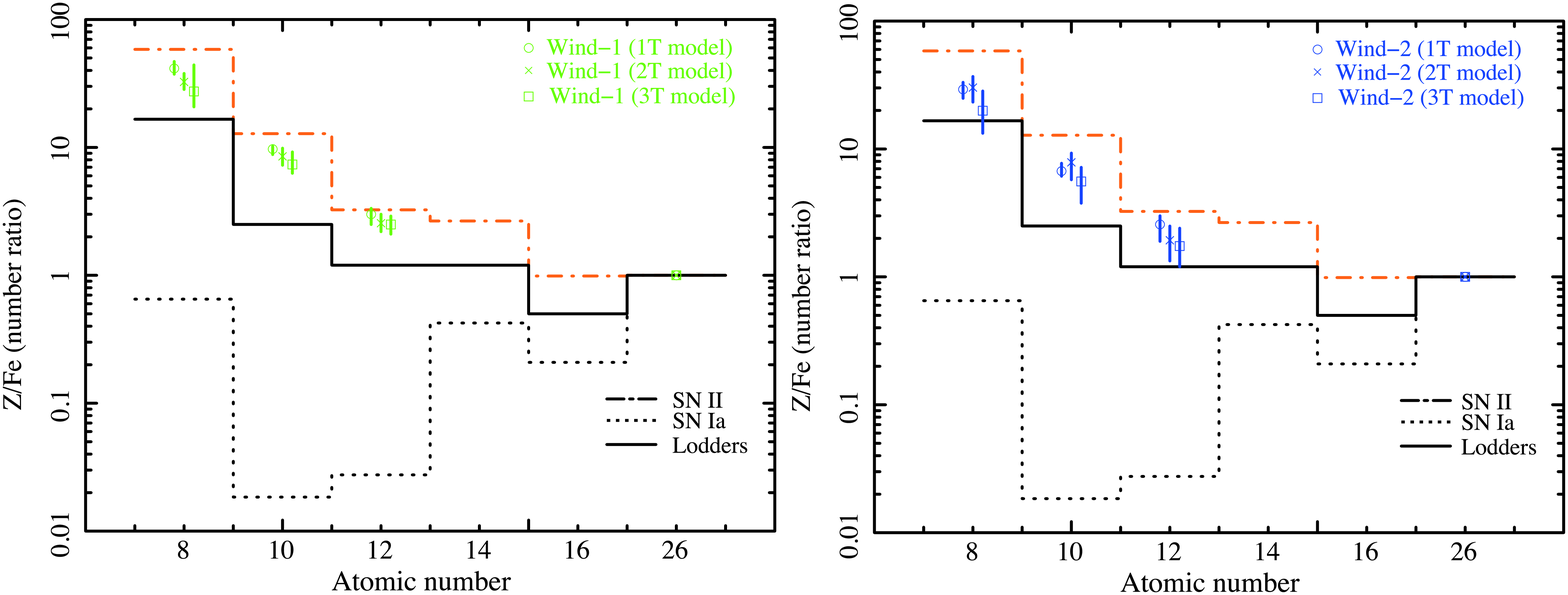}}
\caption{
Abundance ratios of O, Ne, and Mg to Fe for the 1T, 2T, and 3T model of the M~82 Wind-1 (left) 
and Wind-2 (right) regions. 
The abundance patterns of Wind-1 and Wind-2 region are also shown. 
Solid, dot-dashed, and dotted lines represent the number ratios of metals to 
Fe for the solar abundance, for SN II, and for SN Ia products \citep{lodders_03, iwamoto_99, nomoto_06}, respectively. 
}\label{figure6}
\end{center}
\end{figure*}

To investigate the sensitivity of metal abundance determination on the ISM temperature, 
we also employed the 1T and 3T models and fitted the spectrum again, assuming 
that the Disk contamination level is 7\%. 
The 1T and 3T model fits result in ${\chi^2}$/d.o.f.=479/296 and 351/292, respectively.
In particular, 1T model gives a much worse fit to the data as compared to the 2T model 
which had a ${\chi^2}$/d.o.f.=354/294. 
The derived parameters are summarized in table \ref{table3}. 
Furthermore, we calculated confidence contours between the metal abundances 
(O, Ne, and Mg) relative to Fe, using the 1T, 2T, and 3T ISM models.  
From these contours, we derived the abundance patterns (table \ref{table3}), 
which have smaller uncertainties as compared to the absolute abundance values.  
Figure \ref{figure6} shows the abundance patterns of M~82 Wind-1 with 1T, 2T, and 3T models. 
These metal abundance patterns are consistent within error in all models, 
although the 3T model shows the largest uncertainties.

\begin{table*}
\caption{
Summary of the best-fit parameters for the Wind-1 region with 1T, 2T, and 3T for ISM models.}
\label{table3}
\begin{center}
\begin{tabular}{lccccccc} \hline \hline
Parameters & & Wind-1 &  &       \\ 
           &  & 1T for ISM & 2T for ISM &   3T for ISM   \\
           &  &  Disk 7\%  &  &                \\\hline
$N_{\rm H_{G}}$ & (10$^{20}$ cm$^{-2}$) & 4.0 (fix) & 4.0 (fix) & 4.0 (fix)  \\
$kT_{\rm 1T}$ & (keV) &  0.38$\pm0.01$ & 0.26$\pm0.01$ & 0.26$\pm0.01$  \\ 
$kT_{\rm 2T}$ & (keV) & - & 0.57$\pm0.02$ & 0.54$^{+0.04}_{-0.05}$    \\ 
$kT_{\rm 3T}$ & (keV) & - & - & 0.97$^{+1.04}_{-0.21}$    \\ 
O & (solar) & 0.88$^{+0.10}_{-0.09}$  & 1.00$^{+0.17}_{-0.12}$ & 0.99$^{+0.16}_{-0.13}$   \\
Ne & (solar) & 1.29$^{+0.14}_{-0.13}$ & 1.73$^{+0.29}_{-0.22}$  & 1.75$^{+0.32}_{-0.26}$   \\
Mg, Al & (solar) & 0.85$^{+0.14}_{-0.13}$ & 1.09$^{+0.23}_{-0.18}$ & 1.25$^{+0.32}_{-0.24}$   \\
Fe, Ni & (solar) & 0.35$\pm0.04$ &  0.51$^{+0.08}_{-0.06}$ & 0.60$^{+0.14}_{-0.12}$     \\
O/Fe & (solar) & 2.5$^{+0.33}_{-0.25}$ & 1.96$^{+0.32}_{-0.25}$ & 1.65$^{+1.01}_{-0.40}$   \\
Ne/Fe & (solar) & 3.86$^{+0.21}_{-0.36}$ & 3.39$^{+0.54}_{-0.50}$ & 2.93$^{+0.74}_{-0.43}$  \\
Mg/Fe & (solar) & 2.5$^{+0.28}_{-0.42}$ & 2.13$^{+0.37}_{-0.30}$ & 2.08$^{+0.34}_{-0.33}$  \\[2.0ex]
${\chi^2}$/d.o.f. & & 479/296 & 354/294 & 351/292  \\ \hline \hline
\end{tabular}
\end{center}
\end{table*}

\subsection{Wind-2 Region}

We fitted the Wind-2 region spectra after subtracting 1\% of the Disk emission 
and the background emission accumulated over three background regions
with the same model as the Wind-1 region.
In this case, the energy ranges are 0.4--5.0 keV for the BI CCD and 
0.5--5.0 keV for the FI CCDs. 
The fit statistic of the 2T model was not good at ${\chi^2}$/d.o.f.=397/292. 
Because positive residuals exist above 2~keV, 
we attempted setting the contamination at 1.5\% and 2\% intensity of the
Disk emission. After subtracting and refitting, we found ${\chi^2}$/d.o.f. 
of 360/303 and 668/295, respectively. 
We accepted 1.5\% as the contaminant ratio of Disk emission to Wind-2 region. 
The resultant parameters and fitted spectra are summarized 
in table \ref{table4} and figure \ref{figure5}, respectively.

In Wind-2 region, we also employed the 1T and 3T models in fitting, 
assuming that the Disk contamination level is 1.5\%. 
The fitting of 1T and 3T model gives ${\chi^2}$/d.o.f.=448/305 and 352/301, respectively.
Again, the 1T model returns a worse ${\chi^2}$/d.o.f. of the 2T case 360/303. 
The metal abundance ratios are derived from confidence contours and are 
plotted in figure \ref{figure6}. 
Although the O/Fe ratio of the 3T model is slightly lower than the others, 
there is no large discrepancy amongst the various temperature models.

\begin{table*}
\caption{
Summary of the best-fit parameters for the Wind-2 region with 1T, 2T, and 3T for ISM models.}
\label{table4}
\begin{center}
\begin{tabular}{lccccccc} \hline \hline
Parameters & & Wind-2 &       \\ 
           &  & 1T for ISM & 2T for ISM &   3T for ISM   \\
           &  & Disk 1.5\% & &               \\\hline
$N_{\rm H_{G}}$ & (10$^{20}$ cm$^{-2}$) & 4.0 (fix) & 4.0 (fix) & 4.0 (fix) \\
$kT_{\rm 1T}$ & (keV) & 0.32$\pm0.01$ & 0.24$\pm0.01$ & 0.24$\pm0.01$  \\ 
$kT_{\rm 2T}$ & (keV) & - & 0.53$^{+0.06}_{-0.05}$ & 0.50$^{+0.08}_{-0.06}$    \\ 
$kT_{\rm 3T}$ & (keV) & - & - & 1.18$^{+0.47}_{-0.25}$   \\ 
O & (solar) & 0.61$^{+0.09}_{-0.07}$ & 0.69$^{+0.12}_{-0.10}$ & 0.69$^{+0.13}_{-0.10}$    \\
Ne & (solar) & 0.94$^{+0.15}_{-0.12}$  & 1.20$^{+0.23}_{-0.19}$ & 1.28$^{+0.31}_{-0.25}$    \\
Mg, Al & (solar) & 0.70$^{+0.20}_{-0.17}$ & 0.62$^{+0.21}_{-0.17}$ & 0.84$^{+0.38}_{-0.27}$  \\
Fe, Ni & (solar) & 0.34$^{+0.06}_{-0.05}$ & 0.38$^{+0.08}_{-0.06}$ & 0.57$^{+0.21}_{-0.15}$    \\
O/Fe & (solar) & 1.76$^{+0.24}_{-0.26}$ & 1.82$^{+0.40}_{-0.42}$ & 1.2$^{+0.51}_{-0.40}$ \\
Ne/Fe & (solar) & 2.67$^{+0.41}_{-0.23}$ & 3.13$^{+0.57}_{-0.84}$ & 2.22$^{+0.64}_{-0.72}$ \\
Mg/Fe & (solar) & 2.14$^{+0.36}_{-0.56}$ & 1.61$^{+0.47}_{-0.50}$ & 1.45$^{+0.55}_{-0.45}$ \\[2.0ex]
${\chi^2}$/d.o.f. & & 448/305 & 360/303 & 352/301 \\ \hline \hline
\end{tabular}
\end{center}
\end{table*}

\subsection{Wind-3 Region}

\subsubsection{XMM-Newton Analysis of Source~A Spectra}
\label{agn}
Before fitting the Wind-3 spectra, we estimate the emission from Source~A based upon XMM-Newton data. 
We extracted source spectra within a 15$''$ radius circular aperture centered at 
($\alpha$, $\delta$)=(\timeform{09h55m14s}, \timeform{+69D47'35''})\@.
The background spectra were extracted from an annular ring  
and subtracted from Source~A.
We extracted the spectra of MOS~1, MOS~2, and PN and fitted these simultaneously 
in the 0.3--7.0 keV energy range with the following model;
phabs$_{\rm A}$ $\times$ PL$_{\rm A}$, with all parameters set free.
The best fit values are listed in table \ref{table5}.
The Source~A spectra are reproduced by the absorbed PL model with a photon index of 
1.76$^{+0.10}_{-0.09}$, which is consistent with a typical 
active galactic nucleus spectrum.

\begin{table}
\caption{
Best-fit parameters for Source~A with phabs$_{\rm A}$ $\times$ PL$_{\rm A}$}
\label{table5}
\begin{center}
\begin{tabular}{lcccc} \hline\hline
Parameters & &    \\ \hline
$N_{\rm H_{A}}$ & (10$^{20}$ cm$^{-2}$) & 11.2$^{+2.20}_{-2.44}$  \\
$\Gamma_{\rm A}$& & 1.76$^{+0.10}_{-0.09}$ \\ 
Source~A flux$^{\ast}$ & (erg cm$^{-2}$ s$^{-1}$) & 2.2$^{+0.3}_{-0.2}$$\times$10$^{-13}$  \\ 
${\chi^2}$/d.o.f. & & 167/178  \\ \hline \hline
\end{tabular}
\end{center}
\parbox{\textwidth}{\footnotesize
\footnotemark[$\ast$]
Flux within the accumulated region between 0.5 and 8 keV.
}
\end{table}

\subsubsection{Estimation of Background Emission}
\label{bgd}
As a result of lower surface brightness of the Wind-3 region as compared to Wind-1 and Wind-2 (figure \ref{figure3}), we decided to perform simultaneous fitting of the Wind-3 and background spectra. 
Background spectra were extracted from three regions. 
Because the NXB is subtracted, 
the background components consist of Cosmic X-ray Background (CXB) and 
Galactic thermal emission.
We represented CXB with a PL model and Galactic thermal emission 
as a two-temperature apec model \citep{smith_01}.
Empirically, 
one thermal component represents the sum of solar wind charge exchange
(SWCX) and local hot bubble (LHB),
while Milky Way halo (MWH) emits the other thermal component \citep{yoshino_09}.

We fitted the spectra with the following models; 
phabs$_{\rm G}$ $\times$ (PL$_{\rm CXB}$ + apec$_{\rm MWH}$) + apec$_{\rm LHB}$.
The `phabs$_{\rm G}$' model represents the photoelectric absorption, whose column density 
was fixed to the Galactic value of 4.0 $\times 10^{20} cm^{-2}$ 
in the direction of M~82. 
The `apec$_{\rm MWH}$' and `apec$_{\rm LHB}$' models correspond to thermal emission 
from our Galaxy, 
with the metal abundances fixed at the solar value and zero redshift. 
The spectra from the BI and FI CCDs were simultaneously fitted in the 0.4--5.0 and 
0.5--5.0 keV range, respectively.
In table \ref{table6}, we list the derived parameters, which are consistent 
with \citet{tsuru_07}.
This model reproduced the spectra approximately, with ${\chi^2}$/d.o.f.=312/255.
The derived photon index of PL$_{\rm CXB}$ is 1.43$^{+0.09}_{-0.11}$, 
which is consistent with \citet{kushino_02}. 
Although the temperature of apec$_{\rm MWH}$ model is higher than that of the common 
MWH component ($\sim$0.25~keV), 
there are known to be high temperature Galactic 
component ($\sim$0.6~keV) present in some portions of the sky \citep{yoshino_09}. 
Furthermore, similar results are reported in \citet{sato_08} and \citet{konami_10}. 
Since the Galactic emission is considered to have spatial dependencies, 
we regard this model to be appropriate for estimating the background emission in this analysis.

\begin{table}
\caption{
Best-fit parameters for the background regions with  apec components + PL model.$^{\ast}$}
\label{table6}
\begin{center}
\begin{tabular}{lccc} \hline\hline
Parameters               &             &                          \\ \hline
$N_{\rm H_{G}}$               & (10$^{20}$ cm$^{-2}$) & 4.0 (fix)   \\
$\Gamma_{\rm CXB}$        &             & 1.43$_{-0.11}^{+0.09}$            \\ 
 [1.0ex]
$kT_{\rm MWH}$            & (keV)       & 0.70$\pm{0.07}$        \\ 
Abundance                & (solar)     & 1 (fix)                    \\ 
$kT_{\rm LHB}$             & (keV)       & 0.21$_{-0.02}^{+0.03}$         \\ 
Abundance                 & (solar)     & 1 (fix)                   \\ 
$Norm_{\rm LHB}/Norm_{\rm MWH}$  &             & 0.97$_{-0.19}^{+0.12}$        \\
[2.0ex] 
${\chi^2}$/d.o.f.         &             & 312/255                   \\ \hline \hline
\end{tabular}
\end{center}
\parbox{\textwidth}{\footnotesize
\footnotemark[$\ast$]
The apec components for spectra in the background region of \\
M~82 with absorbed MWH component, LHB component for \\the Galactic emission, 
and a PL model for CXB.}
\end{table}

\subsubsection{Simultaneously Fitting of Wind-3 and Background Regions}
\label{halo3}

We fitted the Wind-3 spectra and background simultaneously with following models;
phabs$_{\rm G}$ $\times$ vapec$_{\rm 1T, 2T}$ + phabs$_{\rm A}$ $\times$ PL$_{\rm A}$ 
+ background models.
Here, we used energies ranging from 0.4 (BI) or 0.5 (FI) to 5.0 and 8.0 keV for the 
background and Wind-3 regions, respectively.
The first and second terms represent the outflow
and Source-A, respectively.
The photon index of the PL model for CXB was fixed to 
1.43, which was best fit parameter in section \ref{bgd}.
All the other background parameters were linked between the two regions.

\begin{figure*}
\begin{center}
\centerline{
\FigureFile(\textwidth,\textwidth){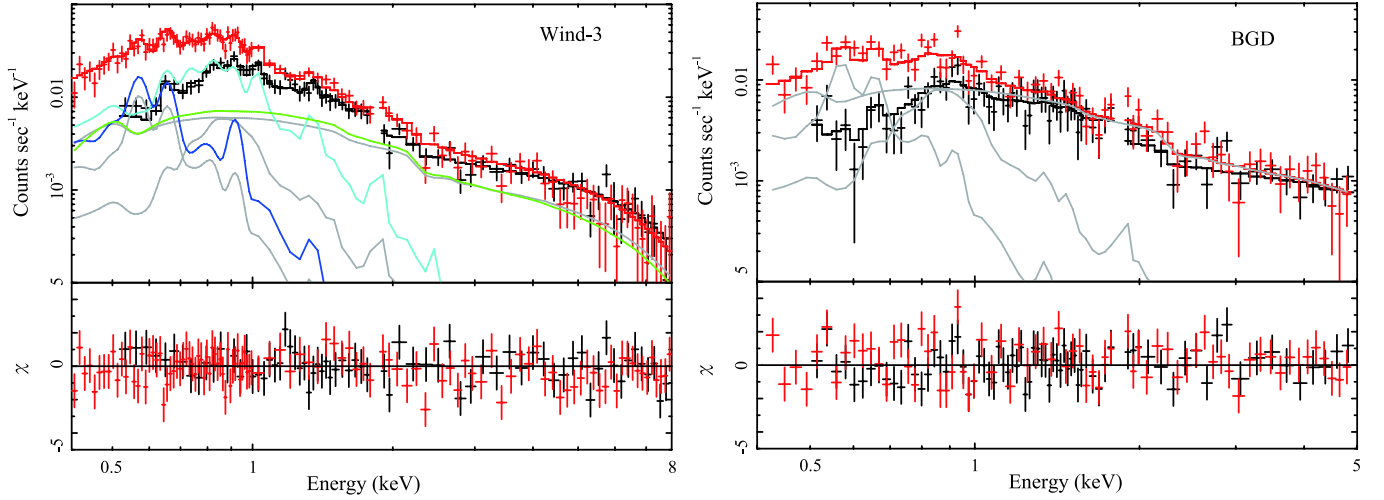}}
\caption{
NXB-subtracted XIS0 (black) and XIS1 (red) spectra 
of M82 (left panel), and those of the background region (right panel), 
shown without removal of instrumental response. 
The left panel employs the two-temperature model for the ISM.
Black and red lines show the best-fit model for XIS0 and XIS1, 
respectively. For simplicity, only the model components for XIS1 spectra 
are shown. Blue and light blue lines are the ISM components, 
the green line is emission from Source~A modelled as an absorbed power-law (PL$_{\rm A}$), 
gray lines are the Galactic background emission (apec$_{\rm MWH}$ and apec$_{\rm LHB}$) 
and the CXB components, respectively. 
The background components are common between 
the on-source and background spectra, but scaled to the respective 
data accumulation area.
}\label{figure7}
\end{center}
\end{figure*}

We employed the 1T and 2T models in fitting, and found ${\chi^2}$/d.o.f. $=$ 
601/543 and 587/541, respectively. 
The fitted spectra and derived parameters are shown in figure \ref{figure7} 
and table \ref{table7}, respectively.
The temperatures and metal abundances between the 1T and 2T models
are consistent with each other, although the 2T model allows higher values of abundances.
When employing the 3T model, the third temperature could not be constrained 
and the normalization is less than 10\% of the those of other components. 
The spectra of Wind-3 region do not require a third temperature component.

\begin{table*}
\caption{
Summary of the best-fit parameters for the Wind-3 region with 1T or 2T for ISM models.}
\label{table7}
\begin{center}
\begin{tabular}{lccc} \hline \hline
Parameters & &  Wind-3 &     \\ 
           &  &  1T for ISM & 2T for ISM \\\hline
$N_{\rm H_{G}}$ & (10$^{20}$ cm$^{-2}$) & 4.0 (fix) & 4.0 (fix)   \\
$kT_{\rm 1T}$ & (keV) &  0.41$^{+0.05}_{-0.03}$ &  0.20$^{+0.04}_{-0.06}$ \\ 
$kT_{\rm 2T}$ & (keV) &  - & 0.49$^{+0.08}_{-0.09}$   \\ 
$Norm_{\rm 1T}/Norm_{\rm 2T}$ & &  - & 0.53$^{+0.72}_{-0.35}$  \\ 
O & (solar) & 0.57$^{+0.27}_{-0.17}$ & 0.68$^{+0.55}_{-0.26}$  \\
Ne & (solar) & 0.95$^{+0.44}_{-0.24}$ & 1.48$^{+1.21}_{-0.47}$  \\
Mg, Al & (solar) & 0.61$^{+0.25}_{-0.23}$ & 0.90$^{+0.64}_{-0.34}$  \\
Fe, Ni & (solar) & 0.21$^{+0.12}_{-0.07}$  & 0.31$^{+0.31}_{-0.07}$   \\
O/Fe & (solar) & 2.73$^{+1.66}_{-0.98}$ & 2.17$^{+0.77}_{-0.97}$  \\
Ne/Fe & (solar) &  4.55$^{+1.70}_{-1.30}$ & 4.84$^{+1.98}_{-1.46}$ \\
Mg/Fe & (solar) &  2.90$^{+1.39}_{-1.15}$ & 2.90$^{+1.10}_{-1.15}$   \\[2.0ex]
$N_{\rm H_{A}}$ & (10$^{20}$ cm$^{-2}$) &  11.2 (fix) &  11.2 (fix)  \\
$\Gamma_{\rm A}$& & 1.73$^{+0.23}_{-0.16}$ & 1.68$^{+0.23}_{-0.16}$ \\ 
Source~A flux$^{\ast}$ & (erg cm$^{-2}$ s$^{-1}$) &  2.1$^{+0.5}_{-0.2}$$\times$10$^{-13}$ & 2.1$^{+0.61}_{-0.39}$$\times$10$^{-13}$ \\[2.0ex] 
${\chi^2}$/d.o.f. & & 601/543 & 587/541 \\ \hline \hline
\end{tabular}
\end{center}
\parbox{\textwidth}{\footnotesize
\footnotemark[$\ast$]
Flux within the accumulated region between 0.5 and 8 keV.} 
\end{table*}

\subsection{Search for CX emission in Wind regions}
\label{search_cx}

As seen in section \ref{disk spectra}, the CX emission may contribute in addition to thermal emission in Disk region. 
To search for CX emission in Wind regions, we fitted the spectra of Wind-1, Wind-2, and Wind-3 regions with 
a 2T thermal model including six Gaussians representing O and Ne emission lines, by setting O and Ne abundance of 
vapec models to zero, as described in section \ref{disk spectra}. 
The ratios of (O \emissiontype{VII} K$\beta$ + O \emissiontype{\rm VIII} Ly$\alpha$)/O \emissiontype{VII} K$\alpha$, 
O \emissiontype{\rm VIII} Ly$\beta$/(O \emissiontype{VII} K$\beta$ + O \emissiontype{\rm VIII} Ly$\alpha$), 
(Ne \emissiontype{IX} K$\beta$ + Ne \emissiontype{X} Ly$\alpha$)/Ne \emissiontype{IX} K$\alpha$, and
Ne \emissiontype{X} Ly$\beta$/(Ne \emissiontype{IX} K$\beta$ + Ne \emissiontype{X} Ly$\alpha$) are derived 
in Disk, Wind-1, Wind-2, and Wind-3 regions.

\section{Discussion}
\label{sec:discuss}

\subsection{Excess emission of Ly$\beta$ lines of H-like O and Ne}
\label{cx_o}

\begin{figure*}
\begin{center}
\centerline{
\FigureFile(\textwidth,\textwidth){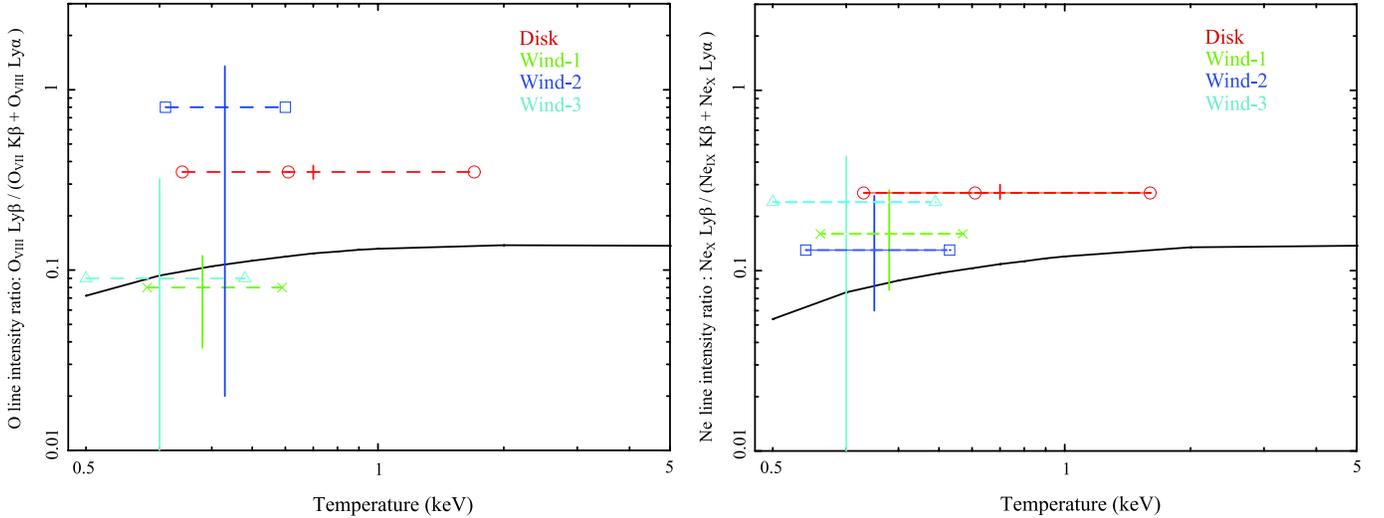}}
\caption{
Left; The red, green, blue, and light blue data represent the values of the line intensity ratio 
O \emissiontype{\rm VIII} Ly$\beta$/(O \emissiontype{VII} K$\beta$ + O \emissiontype{\rm VIII} Ly$\alpha$) vs. 
the thermal model temperatures, for the Disk, Wind-1, Wind-2, and Wind-3 regions, respectively. 
The temperatures used are those determined from thermal models including six Gaussians representing O and Ne emission lines, 
by setting O and Ne abundance of vapec models to zero.
The solid line is same ratio predicted for a CIE plasma using the vapec code \citep{smith_01}.
Right; Symbols and curves are as in the left-hand panel, but for the line ratio  
Ne \emissiontype{X} Ly$\beta$/(Ne \emissiontype{IX} K$\beta$ + Ne \emissiontype{X} Ly$\alpha$).
}\label{figure8}
\end{center}
\end{figure*}

There are residual structures at 0.5--0.7 keV and $\sim$ 1.2 keV
in the spectra of the Disk region fitted with the thermal models.
These residuals are well fitted by excess emission of Ly$\beta$ lines
of O \emissiontype{\rm VIII} and Ne \emissiontype{\rm X}.
A plausible explanation for these residuals is CX between highly ionized
O/Ne and neutral H/He,
since the CX process gives higher ratios of K$\beta$/K$\alpha$ or K$\gamma$/K$\beta$ 
than those of CIE plasma. 
We calculated confidence contours between the normalizations of the Gaussian lines, 
O \emissiontype{\rm VIII} Ly$\beta$/(O \emissiontype{VII} K$\beta$ + O \emissiontype{\rm VIII} Ly$\alpha$) 
and Ne \emissiontype{X} Ly$\beta$/(Ne \emissiontype{IX} K$\beta$ + Ne \emissiontype{X} Ly$\alpha$), 
using thermal models including six Gaussians representing O and Ne emission lines 
(see subsection \ref{disk spectra}). 
In figure \ref{figure8},
the derived these values are plotted against plasma temperature with
those of the CIE plasma from the APEC code.
Both these ratios for H-like O and Ne in the Disk region have values of $\sim$0.3, 
significantly larger than those predicted for the CIE plasma.
Since the temperature dependence of Ly$\beta$ to Ly$\alpha$ ratio is relatively small,
no combination of temperature components alone can give the observed high Ly$\beta$ excess. 
In contrast, the Wind regions have ratios of O \emissiontype{\rm VIII} Ly$\beta$/(O \emissiontype{VII} K$\beta$ + O \emissiontype{\rm VIII} Ly$\alpha$) and Ne \emissiontype{X} Ly$\beta$/(Ne \emissiontype{IX} K$\beta$ + Ne \emissiontype{X} Ly$\alpha$) which are consistent with those for the CIE plasma.
Especially, in the Wind-1 region, O \emissiontype{\rm VIII} Ly$\beta$/(O \emissiontype{VII} K$\beta$ + O \emissiontype{\rm VIII} Ly$\alpha$) is factor of 3--10 smaller than those in the Disk region.
Therefore, this line ratio analysis 
 indicates that the CX process gives higher Ly$\beta$ to Ly$\alpha$ ratio of H-like O and Ne in the Disk region.

For He-like ions, 
the values of (O \emissiontype{VII} K$\beta$ + O \emissiontype{\rm VIII} Ly$\alpha$)/O \emissiontype{VII} K$\alpha$ and (Ne \emissiontype{IX} K$\beta$ + Ne \emissiontype{X} Ly$\alpha$)/Ne \emissiontype{IX} K$\alpha$ in the Disk and the all Wind regions are consistent with those of CIE plasma.
The cross-sections of interaction of highly ionized H-like ions with neutral H/He atoms are larger than those of He-like ions (\cite{koutroumpa_06} and references therein).
Since K$\beta$ line energies of He-like O and Ne are close to those of
K$\alpha$ lines of H-like ions of the same element,
and since the ratios of He-like to H-like ion K$\alpha$ lines have
very steep temperature dependencies,
K$\beta$ to K$\alpha$ ratios of He-like ions are not suitable for the study of the contribution of CX emission
using CCD spectra.

The Ly$\beta$ lines of H-like O and Ne lie close the expected energies of strong Fe-L lines. 
Large residuals at $\sim$ 1.2 keV have been observed in various kinds of objects (e.g. \cite{yamaguchi_10}).
In elliptical galaxies with hot ISM of $\sim$ 0.6 keV,
residuals at $\sim$ 0.8 keV are also observed (\cite{matsushita_07}; \cite{tawara_08}). 
These residuals have been interpreted as a problem in the Fe-L atomic data (e.g. \cite{brickhouse_00}).
However, there is a clear difference in the residual structures
between the Disk and the Wind-1 regions of M~82.
In these two regions, the O and Ne lines mostly come from the same two temperature components,
0.2--0.3 keV and $\sim$ 0.6 keV, although the ratio of the normalization is different.
Therefore, this spatial variation of the residual structure indicates the difference to be a result of CX rather than uncertainties in the Fe-L atomic data.

\citet{liu_11} discovered excess emission of the forbidden lines in the triplet of the He-like
O and Ne, which are robust evidence of CX. 
They assumed the temperature of thermal emission is $\sim$0.6~keV, which is derived using the ratio of 
Mg \emissiontype{\rm XII} Ly$\alpha$/Mg \emissiontype{\rm XI} K$\alpha$. 
Therefore, the CX emission of O \emissiontype{\rm VII} is a result of the capture of two electrons 
by O \emissiontype{\rm IX} ions, while the cross-section of this process is less than 20\% of that of 
single-electron capture via the CX process of O \emissiontype{\rm VIII}, because the ion fraction of 
O \emissiontype{\rm IX} is dominant at the $\sim$0.6~keV CIE plasma (\cite{liu_11}, \cite{greenwood_01}).
\citet{liu_11} evaluated the contribution of CX to be $\sim$90\% to the O \emissiontype{\rm VII} triplet, 
assuming the temperature of thermal emission to be $\sim$0.4--0.6~keV.
The contribution of CX depends on the temperature structure, since
 the fraction of resonance, intercombination, and forbidden lines are dependent on the 
plasma temperature \citep{porquet_00}.
The ISM of galaxies is empirically assumed to be composed of a multiple temperature plasma. 
In fact, our results need $\sim$0.2~keV temperature plasma for the disk emission, which accounts for $\sim$30\% 
of the He-like triplet of O.
The existence of the 0.2 keV plasma increases the ratio of
the forbidden to resonance lines, 
but this is still much smaller than the observed values in \citet{liu_11}.
With Astro-H, next Japanese X-ray satellite,
 we will be able to resolve these K$\alpha$ and K$\beta$ lines of H-like and He-like ions,
as well as the triplets of He-like lines, and
investigate CX more clearly.

CX should strongly contribute to the X-ray emission 
in the central regions of starburst galaxies, where abundant cold gas and 
highly ionized ions co-exist and interact.
We have detected plausible evidence of CX emission between 
H-like ions (O and Ne) and neutrals (H and He) in the Disk region.
In the `cap' region, 
the ionized outflow from M~82 are thought to collide with cool ambient gas and enhance
the H$\alpha$ and X-ray emission, and may also cause CX emission
(e.g. \cite{devine_99}, \cite{lehnert_99}). 
An emission line consistent with the
C \emissiontype{VI} transition of n = 4 to 1 at 0.459 keV has been marginally
detected, although it is not statistically significant at
the 99\% confidence level \citep{tsuru_07}.

\subsection{Metal Abundance Patterns of All Wind regions}

\begin{figure*}
\begin{center}
\centerline{
\FigureFile(0.7\textwidth,0.7\textwidth){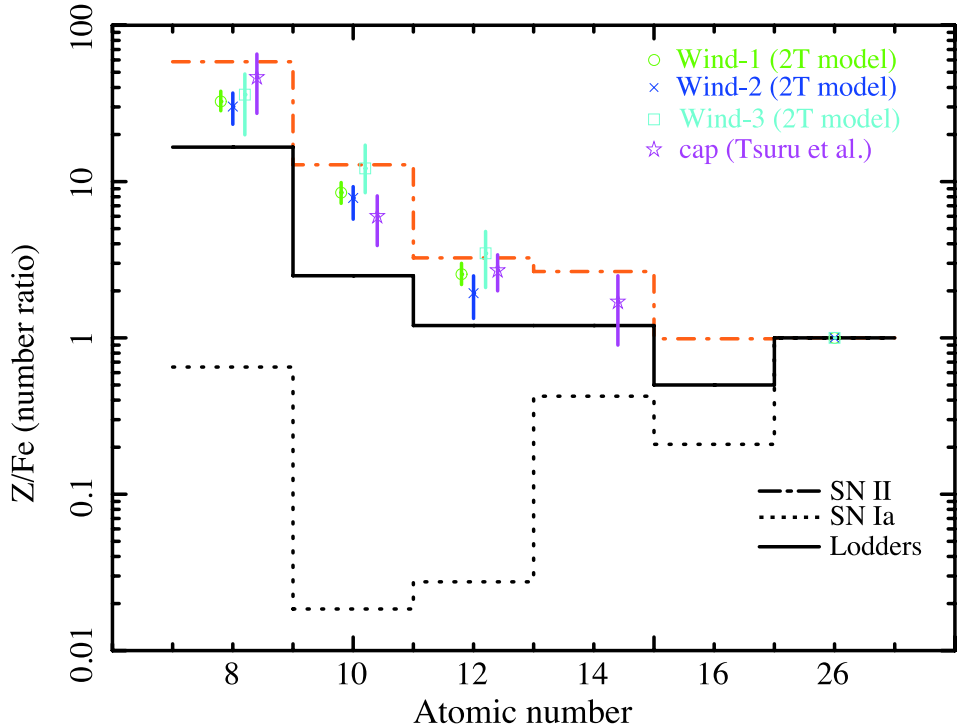}}
\caption{
Abundance ratios of O, Ne, and Mg to Fe for the best-fit model of the M~82 Wind regions. 
The abundance patterns of Wind-1 (green), Wind-2 (blue), Wind-3 (light blue), 
and cap (magenta; \cite{tsuru_07}) regions are also shown. 
Solid, dot-dashed, and dotted lines represent the number ratios of metals to 
Fe for the solar abundance, for SN II, and for SN Ia products \citep{lodders_03, iwamoto_99, nomoto_06}, respectively. 
}\label{figure9}
\end{center}
\end{figure*}

We have successfully derived metal abundances of O, Ne, Mg, and Fe over the entire Wind region of M~82 
for the first time. 
The spectra are well reproduced with thermal models and without evidence of CX emission in the outflow regions. 
In these regions, the amount of neutral gas may be smaller than those in 
the Disk and `cap' regions, and therefore, the effect of the CX may not be important.
As a result, the metal abundances in the Wind regions 
should be more reliable than those in the Disk and `cap' regions.
Furthermore, there no discrepancies among the metal abundance ratios determined using either 
the 1T, 2T, or 3T ISM model over all the Wind regions. 
We calculated confidence contours between the abundance of 
metals (O, Ne, and Mg) relative to Fe, using the 2T model for ISM of all Wind regions.  
From these contours, we derived the abundance patterns 
(table \ref{table3}, table \ref{table4}, and table \ref{table7}), 
which have smaller uncertainties as compared to the absolute abundance values.  
Figure \ref{figure9} shows the abundance patterns of M~82 Wind-1, Wind-2, and Wind-3 within the  
`cap' region \citep{tsuru_07}. 
The distance of these regions from the galactic center range from 2~kpc to 10~kpc.
The O/Fe, Ne/Fe, and Mg/Fe ratios are consistent with no spatial variation.

The calculated SN II and SN Ia yields are also plotted in figure \ref{figure9}. 
The SN II yields by \citet{nomoto_06} refer to an average over the Salpeter initial 
mass function of stellar mass from 10 to 50 $M_{\odot}$, with a progenitor 
metallicity of $Z=0.02$\@.
The SN Ia yields were taken from the W7 model \citet{iwamoto_99}.
The O/Fe, Ne/Fe, and Mg/Fe ratios are a factor of 2--3 higher than the solar values,
and therefore, the outflow of M~82 is significant enriched by SN II yields.
The outflow wind of another starburst galaxy, NGC~4631, observed with Suzaku also shows
enhancements of $\alpha$-elements relative to the Fe abundance ratios \citep{yamasaki_09},
although
its disk region possesses a hot ISM with an approximately solar abundance pattern.
The abundance pattern of the hot ISM in NGC 4258, a spiral galaxy without starburst activity,
is also consistent with the solar ratio \citep{konami_09}. 
These results indicate that starburst galaxies eject metals synthesized by SN II into the intergalactic medium via outflows. 
The derived abundance pattern of the Disk region, especially the very low O abundance relative to the other elements,
is similar to those observed with ASCA and XMM \citep{tsuru_97, ranalli_08}.
However, the observed excess emission of Ly$\beta$ lines and forbidden lines in He-like triplet
indicates that 
a significant fraction of these lines are caused by CX process.
Therefore, it is very difficult to isolate the thermal emission 
owing to the multiple temperature ISM plasma. 
Furthermore, the derived value of $N_H$, $\sim 30\times 10^{20}\rm{cm^{-2}}$,
is significantly higher than the Galactic value of $\sim 4\times 10^{20}\rm{cm^{-2}}$,
whereas those of the Wind regions are consistent to the Galactic value.
The observed value of $N_H$ toward the Disk region means that 85\% of photons in the 
 Ly$\alpha$ line of H-like O must be absorbed, while only 20-30\% of K lines of Mg absorbed. 
The cold gas which absorbs the X-ray emission from the Disk should be non-uniform,
and X-ray photons should be partially absorbed.
Therefore, the observed very low abundance of O may be caused by incorrect modelling of
spectral components of this very complicated plasma in the Disk region.

\section{Conclusion}

We have performed X-ray spectral analysis of the outflow region of M~82 observed with Suzaku. 
Suzaku XIS has a high spectral sensitivity to faint
O lines thanks  to a very small low-pulse-height tail below $\sim$1~keV. 
The spectra from the galactic disk
 are reproduced with 3T thermal models and excess emission of Ly$\beta$ lines
of O \emissiontype{\rm VIII} and Ne \emissiontype{\rm X}, which may be caused by
CX process. 
In the outflow region, there is no excess Ly$\beta$ emission, and
the derived O/Fe, Ne/Fe, and Mg/Fe ratios are a factor of 2--3 higher than the solar ratio.
This abundance pattern indicates that starburst activity enriches the intergalactic medium 
 with SN II yields via outflows.

\bigskip
We thank the referee for providing valuable comments.
We gratefully acknowledge all members of the Suzaku hardware and software 
teams and the Science Working Group. 
SK is supported by JSPS Research Fellowship for Young Scientists.

\end{document}